\documentclass[10pt,conference]{IEEEtran}

\usepackage[utf8]{inputenc}
\usepackage{float}
\usepackage{booktabs} 
\usepackage{amsmath}
\usepackage{amssymb}
\usepackage{listings} 
\usepackage{graphicx}
\usepackage{nameref}
\usepackage{multirow}
\usepackage[inline]{enumitem}
\usepackage{array}
\usepackage{multicol}
\usepackage{graphicx}
\usepackage{diagbox}
\usepackage[table,x11names]{xcolor}
\usepackage{listings}
\usepackage{wrapfig}
\usepackage{lscape}
\usepackage{rotating}
\usepackage{epstopdf}
\usepackage{mathtools}
\usepackage{makecell}
 \usepackage[numbers,sort&compress]{natbib} 
\usepackage{tikz}
\usepackage[framemethod=TikZ]{mdframed}
\usepackage[bottom]{footmisc}
\usepackage{caption}
\usepackage{subcaption}
\usepackage[flushleft]{threeparttable}
\usepackage{braket}
\usepackage{hyperref}
\usepackage{cleveref}

\hypersetup{
    colorlinks=true,
    linkcolor=blue,
    urlcolor=blue,
    citecolor=blue,
    linktocpage=false
}

\title{Visually Exploring Software Maintenance Activities}

\author{\IEEEauthorblockN{Stanislav Levin}
\IEEEauthorblockA{The Blavatnik School of Computer Science\\
 Tel Aviv University\\
Tel-Aviv, Israel\\
stas.levin@cs.tau.ac.il}
\and
\IEEEauthorblockN{Amiram Yehudai}
\IEEEauthorblockA{The Blavatnik School of Computer Science\\
 Tel Aviv University\\
Tel-Aviv, Israel\\
amiramy@tau.ac.il}
}

\date{December 2018}

\begin{document}
\bstctlcite{IEEEexample:BSTcontrol}

\maketitle

\thispagestyle{plain}
\pagestyle{plain}

\begin{abstract}

Lehman's Laws teach us that a software system will become progressively less satisfying to its users over time, unless it is continually adapted to meet new needs.
A line of previous works sought to better understand software maintenance by studying how commits can be classified into three main software maintenance activities. \textit{Corrective}: fault fixing; \textit{Perfective}: system improvements; \textit{Adaptive}: new feature introduction. 

In this work we suggest visualizations for exploring software maintenance  activities in both project and individual developer scopes. We demonstrate our approach using a prototype we have built using the Shiny R framework. In addition, we have also published our prototype as an online demo. This demo allows users to explore the maintenance activities of a number of popular open source projects.

We believe that the visualizations we provide can assist practitioners in monitoring and maintaining the health of software projects. In particular, they can be useful for identifying general imbalances, peaks, deeps and other anomalies in projects' and developers' maintenance activities. 


\end{abstract}

\setlength{\abovecaptionskip}{0.2\baselineskip}

\section{Introduction}
\label{sec:intro}

Software maintenance activities are a key aspect of understanding software evolution, and have been a subject of research in numerous works \cite{swanson1976dimensions, meyers1988, lientz1978characteristics, schach2003determining}. One of the research questions studies have been trying to address, is how does one obtain reliable maintenance profiles of software projects. That is, given a software project, we wish to quantify its maintenance activities in a reliable manner. 
\citet{mockus2000identifying} pioneered an approach which relies on the version control system (VCS) and considers the revisions\footnote{A.k.a. "commits", as per the terminology of the Git \cite{GIT} VCS.} as (maintenance) activity boundaries. Each activity, manifested as a revision in the VCS, can therefore be classified according to a taxonomy of maintenance activities (see \Cref{tab:maintenance-activities}). Iterating over the entire VCS history and classifying its revisions would therefor yield a maintenance activity profile.

\begin{table}[h]
\small
\center
\renewcommand{\arraystretch}{1.5}
\caption{Maintenance activities \cite{mockus2000identifying}}
\label{tab:maintenance-activities}
\begin{tabular}{|c|l|}
    \hline
    \rowcolor{lightgray} \textbf{Maintenance Activity} & \multicolumn{1}{|c|}{\textbf{Intent}}\\ \hline
    Corrective & fault fixing \\ \hline
    Perfective & system improvements \\ \hline
    Adaptive & new feature introduction \\ \hline
\end{tabular}
\vspace{-1em}
\end{table}

Over the past two decades, an array of methods has been suggested for classifying commits into maintenance activities \cite{mockus2000identifying, hindle2009automatic, levinIcsme2016, fischer2003populating, sliwerski2005changes, amor2006discriminating}. We indicated in our previous work \cite{levinPromise2017,levin2019software}, that existing \textit{cross-project} classification methods did not go far beyond the 50\% accuracy mark (see also \Cref{tab:current-quality}). In an attempt to bring commit classification into maintenance activities closer to being ``production ready'', we suggested a method that was able to achieve 76\% accuracy and Kappa of 63\% when tested on cross-project commits \cite{levinPromise2017}.

\begin{table}[ht]    
\small    
    \center
    \renewcommand{\arraystretch}{1.5}
    \caption[Commit classification methods]{Commit classification methods, current results \cite{levin2019software}}
    \label{tab:current-quality}
    \begin{threeparttable}
        \begin{tabular}{|l|c|c|c|}        
            \hline
            \rowcolor{lightgray} \multicolumn{1}{|c|}{\textbf{Study}} & \textbf{Scope} &  \textbf{Accuracy} & \textbf{F$_1$}\tnote{*} \\    \hline
            \citet{levinPromise2017} & Cross Project & 76\% &0.76 \\ \hline        
           \centering  \multirow{2}{*}{\citet{hindle2009automatic}} & Single Project & 70\% & 0.69  \\ \cline{2-4}
             & Cross Project &  52\% & 0.51  \\ \hline
            \citet{amor2006discriminating} &  Single Project & 70\% & N/A \\ \hline        
            \citet{mockus2000identifying} & Single Project & 61\% & N/A \\ \hline 
        \end{tabular}    
        \begin{tablenotes}
            \item[*] The F$_1$-measure. A.k.a. the F-measure. \cite{sasaki2007truth}.
        \end{tablenotes}
    \end{threeparttable}
\end{table} 

In this work we build upon the recent efforts to improve classification quality, and seek to utilize the higher quality classification \cite{levinPromise2017} in order to enable the visual exploration of maintenance activities. Despite the fact our visualizations can be applied to any commit classification method, it is important to note that the overall usability of our approach is highly dependent on the reliability of the reported maintenance activities. The latter is dictated by quality of the particular commit classification method used to produce the visualized dataset (see also \Cref{sec:target}).

We demonstrate our approach with a prototype tool we have built, and made available as a public online demo \cite{shiny-maintenance-activities}. The prototype visualizes maintenance activities and provides users with a number of key exploration features. The source code for the prototype has been made public on GitHub \cite{soft-evo-github}.

This paper proceeds as follows. \Cref{sec:related-work} covers previous studies on visualization of software maintenance and evolution. In \Cref{sec:vis-dimentions} we review the features provided as part of our prototype and map them along the principled dimensions framework for software visualization \cite{maletic2002task}.
In \Cref{sec:discussion} we suggest the notion of balanced maintenance activity profiles and discuss how our visualizations can help in identifying such profiles. 
\Cref{sec:future-work} concludes this paper.

\section{Related work}\label{sec:related-work}


\citet{lanza2001evolution} suggested the evolution matrix, which combined software visualization and software metrics in order to deal with the complexity brought about by large amount of data. In the evolution matrix, each column of the matrix represented a version of the software, while each row represented the different versions of the same class. The evolution matrix allowed reasoning on both system and individual class levels.

\citet{german2006visualizing} visualized evolutionary aspects such as file ownership, commit frequency, file coupling \& activity, etc. Their tool, softChange, targeted researchers as the user audience. \citet{german2006visualizing} stressed that it is the software evolutionist who needs to apply experience and insight to explain how the software has evolved, tools merely help in the process. 

\citet{van2004studying} demonstrated how high-level visualization of commits (to the VCS) can be used for recognizing relevant changes in a system's evolution. They plotted file releases against date of release to identify architectural patterns in evolution and team productivity.

In contrast to the evolution matrix \cite{lanza2001evolution} and softChange \cite{german2006visualizing}, which display the evolution of classes and files (respectively), we wish to visualize software maintenance activities (see \Cref{sec:target}).
In this regard our approach is more akin to that of \citet{van2004studying}. However, the underlying data we visualize, and the exploration features we provide, are quite different.

\section{Visualizing Software Maintenance Activities}\label{sec:vis-dimentions}

\citet{maletic2002task} suggested a framework for evaluating and developing software visualization systems. 
This framework included five dimensions:
\begin{itemize}
    \item \textit{Tasks.} Why is the visualization needed?
    \item \textit{Audience.} Who will use the visualization?
    \item \textit{Target.} What is the data source to represent?
    \item \textit{Representation.} How to represent it?
    \item \textit{Medium.} Where to represent the visualization?
\end{itemize}
We find these dimensions helpful for putting our visualizations in context, and address them in detail in this section.

\subsection{Why is the visualization needed?} 

The task we wish to facilitate (or enable) is identifying potential anomalies in the maintenance activity profiles of software projects. 
Software maintenance has long been characterised by its (huge) costs \cite{bennett2000software,pigoski1996practical,seacord2003modernizing,lientz1978characteristics}. Early detection of anomalies may help reduce these costs and prevent escalations. We also hope our visualizations can be used to improve software quality in the long term.

\subsection{Who will use the visualization?} 

Our visualizations may be helpful to users with varying technical skills.
Team managers, project and product managers are the primary roles we believe would be interested in having access to the kind of high-level information we provide. These roles are typically stakeholders in the project's success and are likely to be enthusiastic about early detection of potentially harmful anomalies in maintenance activity profiles. These roles also involve a level of authority, which can be helpful for navigating the project and/or team towards a desired outcome, following the insights they derive based on the visualizations our tool provides.

\subsection{What is the data source to represent?}\label{sec:target}

The aspect of software we seek to visualize is maintenance activities. In the context of our work, maintenance activities are the result mapping VCS commits onto a predefined taxonomy of activity types. The predominant taxonomy for maintenance activities consists of the following three activities: \textit{Corrective}: fault fixing; \textit{Perfective}: system improvements; \textit{Adaptive}: new feature introduction. This taxonomy is widely popular, \cite{swanson1976dimensions,mockus2000identifying,meyers1988,lientz1978characteristics, levinIcsme2016,levinPromise2017,schach2003determining}, and has been used in classification models devised over a number of decades \cite{mockus2000identifying, hindle2009automatic,levinIcsme2016, fischer2003populating, sliwerski2005changes,amor2006discriminating,levinPromise2017}. 
For the purpose of developing and demonstrating our visualizations, we used the classification method suggested in our previous work \cite{levinPromise2017}, where we were able to achieve an accuracy of 76\% and Kappa of 63\% for cross project commit classification. Moreover, high quality cross project classification model is particularly useful when one has to classify and visualize multiple projects, as we did in this work. It allows one to apply a single model to the entire dataset, relieving the need for training multiple per-project models in order to obtain reliable results.

\subsection{How to represent it?} 

\begin{figure*}
    \centering
    \includegraphics[width=0.8\linewidth]{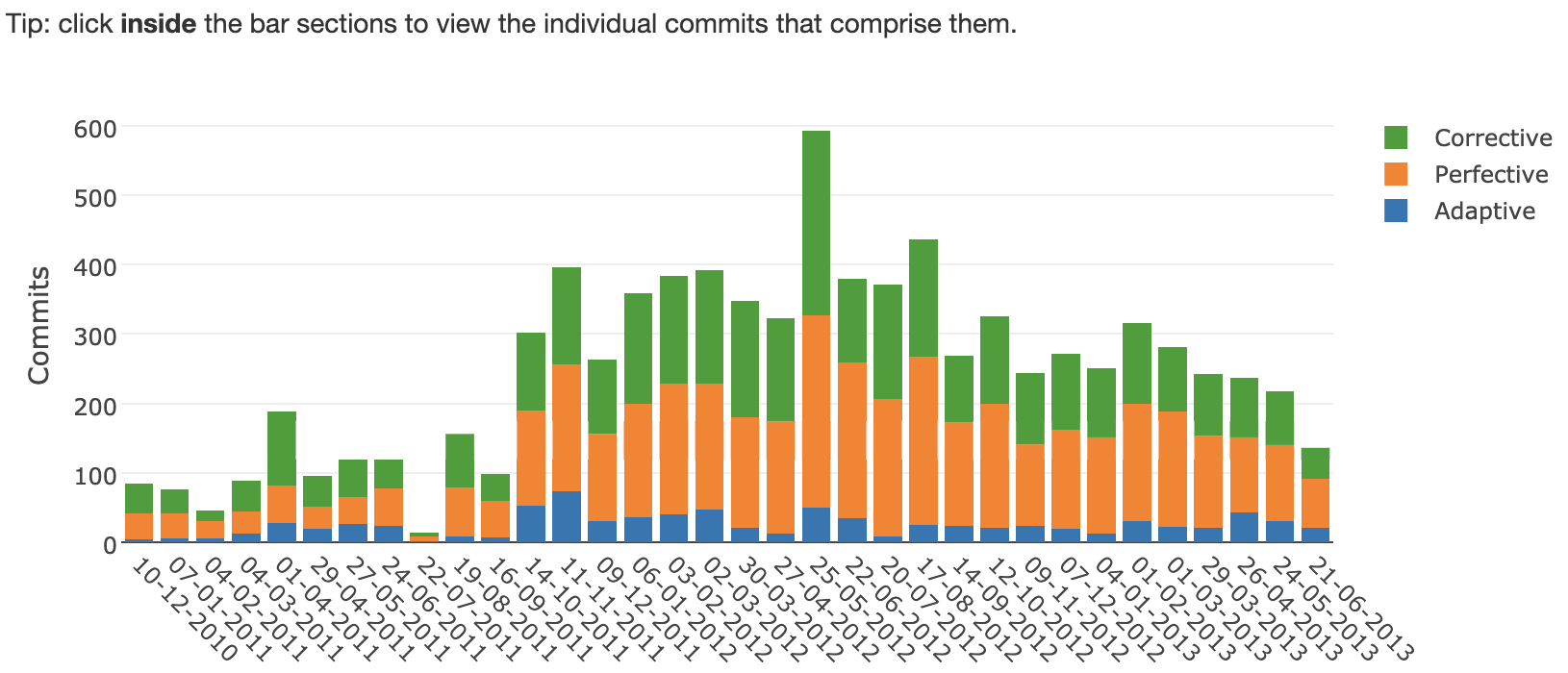}
    \caption{Visualizing maintenance activities}
    \label{fig:bar-diagram}
    \vspace{-1em}
\end{figure*}

Having Few's \cite{few2009now} and Cleveland's \cite{cleveland1985graphical} principles in mind we chose stacked bar diagrams (see \Cref{fig:bar-diagram}) in order to visualize maintenance activities. Each maintenance activity is encoded using a different color, and the three activity types are stacked on top of one another. The x-axis is the time-line, and the y-axis is the activity (commit) count.

Stacked bar diagrams facilitate comparisons between maintenance activities within a given stacked bars column (e.g., what maintenance activity dominated a given time frame), as well as between different stacked bars columns (e.g., which of the time frames had more of a given maintenance activity).
In addition, bar diagrams often allow for an easy detection of anomalies such as peaks and deeps, as well as trends.

As part of the ``Representation'' dimension, \citet{maletic2002task} also included a refined version of the seven principles put forth by \citet{shneiderman2003eyes}. These principles may serve as guidelines for developing navigational needs for visualization applications.
Our projection of these seven principles onto the representation layer of our prototype is as follows.

\paragraph*{Overview}

We provide two main views in our prototype.
Project view, which allows for an exploration of project wide maintenance activities. \\
Developer view, for segmentation of the data by a specific developer within the selected project.
Developer identity can be determined by their name, email, or both. This flexibility can be useful when developers perform commits using a number of different accounts (emails), e.g., when working on an open source project from both their private account and their corporate account.
\paragraph*{Zoom}

\begin{figure}[H]
 \setlength{\belowcaptionskip}{0.5\baselineskip}
  \begin{subfigure}{\linewidth}
  \centering
    \caption{Zooming in on a range of activity buckets}
    \label{fig:zoom}
    \includegraphics[width=0.8\linewidth]{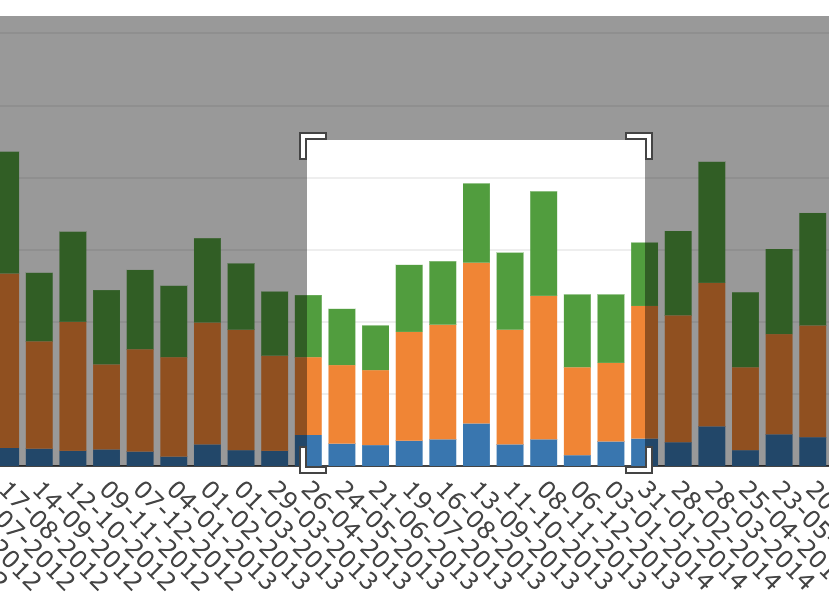}
  \end{subfigure}
  \begin{subfigure}{\linewidth}
  \centering
    \caption{Activity bucket size slider (in days)}
    \label{fig:activity-bucket}
    \includegraphics[width=0.8\linewidth]{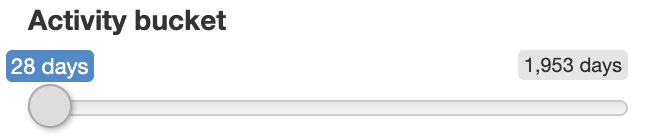}
  \end{subfigure}
  \caption{Configuring zoom and activity bucket size}
\end{figure}

Users can zoom on a specific time period by clicking the left mouse button and dragging the mouse (see \Cref{fig:zoom}).
The zoomed view shows only the maintenance activities that took place during the zoomed-in time period. Zoom can be reset to default by double clicking the mouse.
Another aspect of granularity is the activity bucket size, which determines the stacked bars "column width". Maintenance activities are grouped together (bucketed) according to the specified activity bucket, in days (see \Cref{fig:activity-bucket}).
The default view sets the activity bucket to 28 days to avoid clutter.

\begin{figure}[H]
    \centering
    \includegraphics[width=0.8\linewidth]{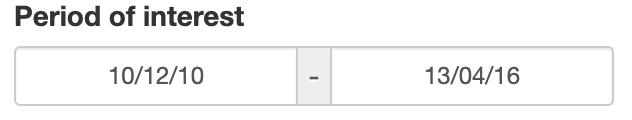}
    \caption{Filtering by period of interest}
    \label{fig:filter-period}
    \vspace{-2em}
\end{figure}

\begin{figure}[H]
    \setlength{\abovecaptionskip}{0.3\baselineskip}
    \centering
    \includegraphics[height=90pt]{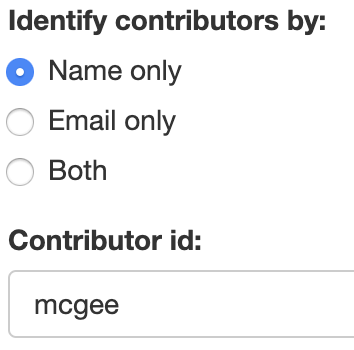}
    \caption{Filtering by developer identifier}    
    \label{fig:dev-id}
\end{figure}

\paragraph*{Filter}
Maintenance activities can be filtered by a number of parameters: project's name and time period (see \Cref{fig:filter-period}).
In the developer centric view, maintenance activities can also be filtered by a developer identifier, which can be a name, an email address, or both (see \Cref{fig:dev-id}).

\begin{figure}[H]
    \centering
    \includegraphics[width=\linewidth]{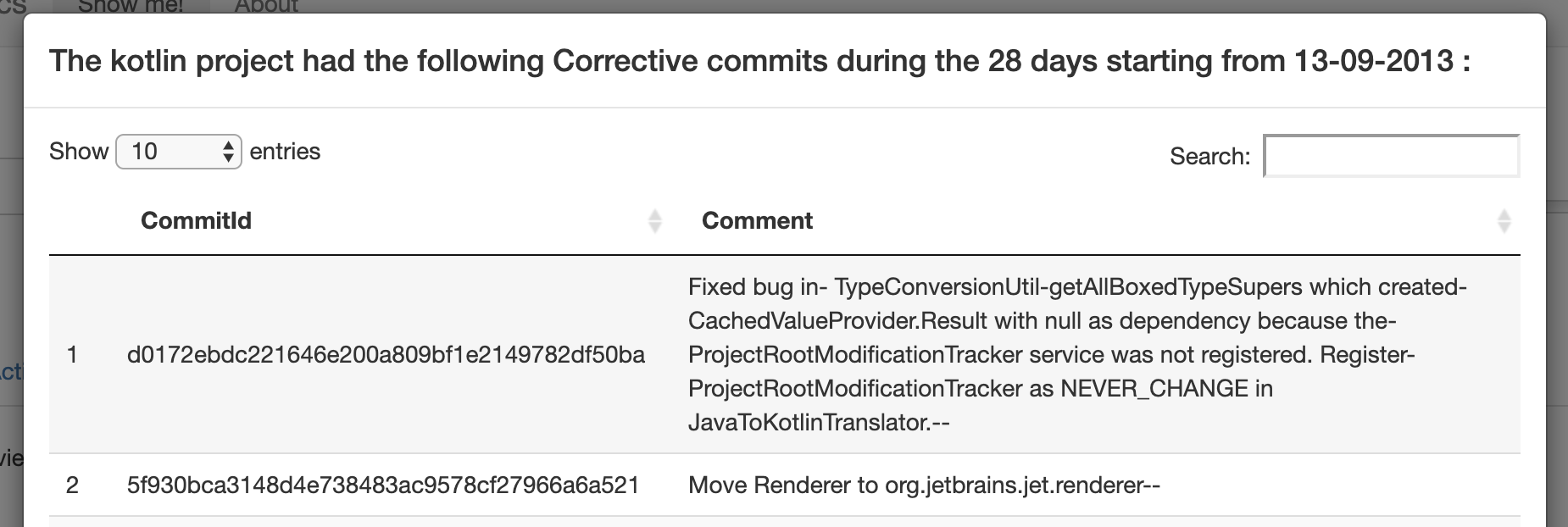}
    \caption{On demand commit level information}    
    \label{fig:details-on-demand}
\end{figure}

\paragraph*{Details-on-demand}
Users can obtain a detailed view of the commits pertaining to a specific maintenance activity and time frame by clicking its color in the corresponding stacked bars column. The detailed view includes commit information such as commit hash and commit message. 
The detailed view can also be searched using free text, in which case it will show any commits where the commit message contained the specified text. In order to avoid overwhelming the users with details, the detailed view shows the first 10 results by default. This number can be configured by the user (see \Cref{fig:details-on-demand}).

\begin{figure}[H]
    \centering
    \includegraphics[width=0.6\linewidth]{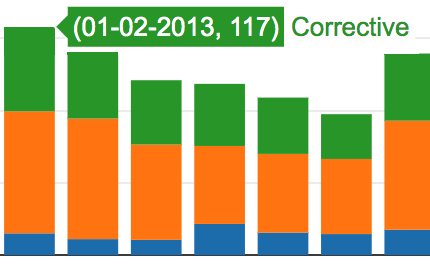}
    \caption{Activity bucket date and count on mouse hover}    
    \label{fig:agg-info}
\end{figure}

\paragraph*{Relate}
The y-axis of the chart is the commits count, since a single commit designated a single activity in our context, the stacked bars can be easily compared to one another. 
The various activity types within the stacked bars can also be compared using their respective color.
By hovering over an area of a given stacked bars column, the corresponding maintenance activity's aggregate information is displayed, see \Cref{fig:agg-info}. This additional numerical information helps in situations where the segmentation within a single stacked bars column is seemingly equal, and visually comparing the areas is not accurate enough.

\paragraph*{History}
As part of the prototype we did not implement any user personalization features such as favorite views or history. Consequently, views are ephemeral and will be reset to default upon closing or refreshing the application window.
Personalization however, could be a great addition to future versions of our tool. For example, while team managers may prefer a more short term view to monitor their team's progress, project or product mangers may want to consider a more long term view of maintenance activities.

\paragraph*{Extract}
Extracting the underlying dataset is possible in a CSV format via the about page, where one can also explore the dataset inline (in the application) without actually downloading it as a file.

\subsection{Where to represent the visualization?} 

Our prototype can be a standalone application, viewed on any laptop or desktop computer which has the Shiny R package and RStudio installed \cite{shiny}.
However, since collaboration is key to the software development process, we believe that it is important to have Software Analytics \cite{buse2010analytics,menzies2013software} tools available online (or over the internal network) so that users can easily share and discuss their (tool based) findings in an effective manner. This was one of the considerations that prompted us to make our prototype available as an online demo, accessible via a standard web browser.

\section{Discussion}\label{sec:discussion}

\begin{figure}
\setlength{\belowcaptionskip}{0.2\baselineskip}
  \begin{subfigure}{0.4\linewidth}
  \centering
    \includegraphics[height=120px]{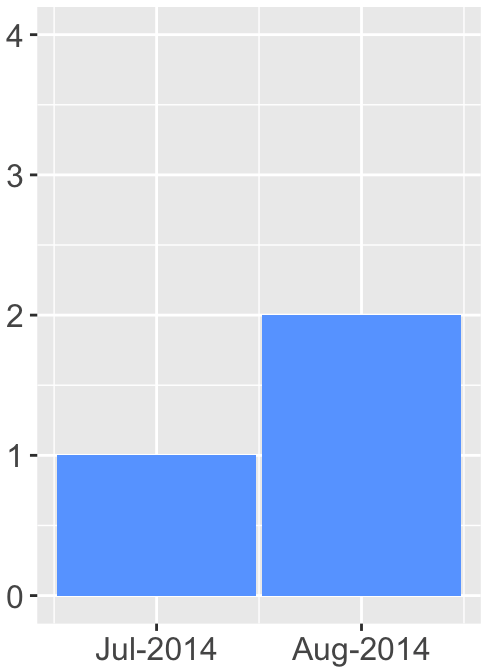}
    \caption{An unbalanced maintenance profile}
    \label{fig:kotlin-only-one}
  \end{subfigure}
  \quad
  \begin{subfigure}{0.4\linewidth}
  \centering
    \includegraphics[height=120px]{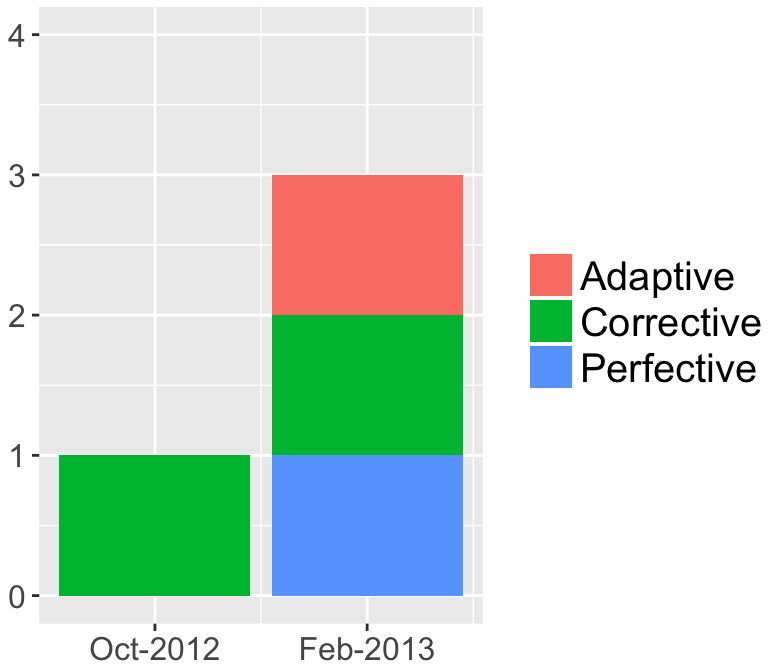}
    \caption{A balanced maintenance profile}
    \label{fig:kotlin-a-c-p}
  \end{subfigure}
  \caption{Two developer profiles from the Kotlin project}
  \vspace{-2em}
\end{figure}

Inspecting the maintenance activities visually, prompted us to consider the balance (see \Cref{fig:kotlin-a-c-p}), or sometimes, the imbalance (see  \Cref{fig:kotlin-only-one}) that may exist in maintenance profiles.
We therefore suggest a notion of a balanced maintenance activity profile, i.e., a profile which includes all three maintenance activity kinds (corrective, perfective, adaptive), and conjecture that it may help developers and teams be more effective and engaged with the project they are working on. 
For example, developers who only (or mostly) engage in corrective activity, may not be exposing themselves to the joy of developing new features (adaptive activity) or making legacy code\footnote{A.k.a., code developers do not like.} look beautiful (perfective activity).

Different project and/or team managers may choose different thresholds for what a balanced (or unbalanced) profile is in terms of proportions. Nonetheless, once these thresholds have been set, profiles that are unbalanced (according to the chosen thresholds) may pose opportunities for continuous improvement. 

This may be of particular interest in the context of open source projects, which tend to heavily rely on community efforts. To that end, well balanced maintenance activity profiles may be something the community needs in order to drive development forward, and ensure that the project gets a fair share of new features, bug fixing, and design improvements - activities that often compete for resources in real-world scenarios.

\section{Conclusions and Future Work}\label{sec:future-work}

In this work we suggest visualizations for exploring software maintenance activities.
We demonstrate our approach in a prototype we have built using the Shiny R package.
In addition, we have also published our prototype as an online demo. This demo allows users to explore the maintenance activities of a number of popular open source projects, providing both project wide and individual developer views.

Using the visualizations we suggest, it is possible to identify unbalanced maintenance activity profiles. We believe that this could be an opportunity to help developers and teams be more effective and engaged with the project they are working on in both commercial and open source development environments.

Future direction may include field studies revolving around adoption and exploring whether our visualizations could appeal to practitioners working on commercial and/or open source projects. It would also be beneficial to learn what real-life tasks \textit{practitioners} believe such visualizations can help with, and/or what changes they would like to suggest to make it more useful for their needs. 

\section*{Acknowledgements}
We thank Dr. Boris Levin for his valuable comments and constructive criticism of the manuscript.

\clearpage

\def\UrlFont{\ttfamily\scriptsize}
\def\UrlBreaks{\do\/\do-}
\bibliography{main}
\bibliographystyle{IEEEtranN}

\end{document}